\documentclass[aps,prb,twocolumn,showpacs]{revtex4}
\usepackage{epsfig}
\usepackage{hyperref}

\begin{document}

\def\il{I_{low}}
\def\iu{I_{up}}
\def\eeq{\end{equation}}
\def\ie{i.e.}
\def\etal{{\it et al. }}
\def\prb{Phys. Rev. {B }}
\def\pra{Phys. Rev. {A }}
\def\prl{Phys. Rev. Lett. }
\def\pla{Phys. Lett. A }
\def\pb{Physica B}
\def\ajp{Am. J. Phys. }
\def\mpl{Mod. Phys. Lett. {B }}
\def\ijmpb{Int. J. Mod. Phys. {B }}
\def\ijp{Ind. J. Phys. }
\def\ijpap{Ind. J. Pure Appl. Phys. }
\def\ibm{IBM J. Res. Dev. }
\def\pjp{Pramana J. Phys.}
\def\pt{Phys. Today}
\def\jpcm{J. Phys. Condensed Matter }

\title{Equilibrium currents in quantum double ring system: A
  non-trivial role of system-reservoir coupling.}  \author{Colin
  Benjamin} \email{colin@iopb.res.in} \affiliation{Institute of
  Physics, Sachivalaya Marg, Bhubaneswar 751 005, Orissa, India}
\author{A. M.  Jayannavar} \email{jayan@iopb.res.in}
\affiliation{Institute of Physics, Sachivalaya Marg, Bhubaneswar 751
  005, Orissa, India}

\date{\today}

\begin{abstract}  
  Amperes law states that the magnetic moment of a ring is given by
  current times the area enclosed. Also from equilibrium statistical
  mechanics it is known that magnetic moment is the derivative of free
  energy with respect to magnetic field. In this work we analyze a
  quantum double ring system interacting with a reservoir. A simple
  S-Matrix model is used for system-reservoir coupling.  We see
  complete agreement between the aforesaid two definitions when
  coupling between system and reservoir is weak, increasing the
  strength of coupling parameter however leads to disagreement between
  the two. Thereby signifying the important role played by the
  coupling parameter in mesoscopic systems.
\end{abstract}


\maketitle 

\section {Introduction} 

Mesoscopics, a fertile branch of Physics deals with systems and
phenomena in the scale of nano- to micrometers. The distinguishing
feature of these systems is that at extremely low temperatures an
electron retains it's phase coherence throughout the sample. These
systems have revealed a new range of unexpected quantum phenomena,
often counter-intuitive\cite{imry,datta}. The notion of the usual
ensemble averaged transport coefficients such as the conductivity,
i.e., local and material specific, has to be replaced by that of
conductance, i.e., global and operationally specific to the sample as
well as the nature of probes of measurement.  Some novel and hitherto
unheard of features in classical physics, e.g., non-local
current-voltage relationships\cite{webb}, breakdown of Ohm's law,
absolute negative resistance (four probe)\cite{butiprl1986},
normal-state Aharonov-Bohm effect\cite{webb}, quantization of
point-contact conductance (Landauer formalism)\cite{vanwees},
universal conductance fluctuations (new form of ergodicity),
persistent currents\cite{BIL}, spin-polarized
transport\cite{prinz,datta1}, coulomb blockade\cite{glazman} and many
novel effects arising due to electron correlations, have been observed
in these systems. Interpretation of these require full recognition of
the wave nature of quasi particles and keeping track of their phase
coherence over the entire sample including the measurement leads and
probes (quantum measurement process). Even the equilibrium properties,
are very sensitive to the nature of statistical ensemble used. The
results differ qualitatively from one ensemble to another\cite{imry}.

\section {Motivation}

Recently it has been proposed that these systems will provide a
testing ground for verifying the violation in the basic laws of
thermodynamics\cite{TA}.  This behavior has been explained by taking
recourse to the effects of entanglement, through which the quantum
system is so interlinked with the bath that the resulting behavior of
the system alone cannot be treated within a conventional thermodynamic
approach. Here the finite coupling between the bath and the system
plays a crucial role. It should be noted that equilibrium
thermodynamics of the super-system comprising of system (sub-system)
plus bath, does not imply standard equilibrium thermodynamics for the
sub-system alone.  In-fact the thermodynamic equilibrium properties of
the system depend on the coupling parameter, unlike equilibrium
statistical mechanics.  For example, it has been shown via an
analytical treatment that the mean orbital magnetic moment, a
thermodynamic property, is determined by the electrical resistivity
(which is related to system-bath coupling parameter) of the
material\cite{dattgup_prl}. What is crucial for dissipative
diamagnetism is that system-bath interaction has to be treated
exactly, there is no clear cut separation between what is the system
and what is the bath- both are inexorably linked to one many-body
system.  In this work, motivated by the above results, we provide a
simple example wherein the equilibrium properties are determined by
the system-reservoir coupling parameter in a {\bf non-trivial manner}.
Our results follow from the consideration of the dephasing of a single
particle quantum coherence while the earlier dynamical treatments
required in addition to quantum coherence, entanglement between system
and bath.  It is in this spirit that we study the persistent current
densities in a quantum double ring system coupled to a reservoir via a
simple voltage probe method due to B\"{u}ttiker\cite{buti}. We
explicitly show that when the coupling parameter is very small there
is perfect agreement between the magnetic moments calculated from the
local currents (via Amperes law) and that from the derivative of free
energy with respect to magnetic field, increasing the strength of
coupling parameter however leads to disagreement between the two.

\section {Background}

It is well known that spontaneous currents which never decay can flow
in super-conducting systems. In 1983, B\"{u}ttiker , Imry and
Landauer\cite{BIL} first predicted that, a normal metal ring threaded
by an Aharonov-Bohm flux $\phi$, in the phase coherent domain carries
persistent currents . These arise because the magnetic flux breaks the
time reversal symmetry thus inducing currents. This is a quantum
effect and the total current flowing in the ring is related to the
derivative of free energy with respect to
flux\cite{cheungibm,cheungprb}. For clean rings it is periodic in flux
with period of $2\pi\phi/\phi_0$. The magnetic flux incidentally plays
the same role as a periodic potential in the Bloch sense, and
therefore one gets the band structure, elastic scattering if present
in the loop induces gaps in the spectrum and reduces the magnitude of
persistent currents. Later in 1985, B\"{u}ttiker \cite{buti}
investigated the effect of a reservoir coupled to a ring.  This simply
means breaking the phase coherence.  Electrons enter the reservoir
lose their phase memory and are re-injected again from the reservoir
with an uncorrelated phase.  The S-Matrix for coupling between system
and reservoir is given by-

\[S_{J}=\left(\begin{array}{ccc}
-(a+b)     & \sqrt{\epsilon}&  \sqrt{\epsilon}\\
 \sqrt{\epsilon}    & a & b\\
\sqrt{\epsilon}      & b  & a
\end{array} \right) \]

wherein, $a=\frac{1}{2}(\sqrt{1-2\epsilon}-1)$ and
$b=\frac{1}{2}(\sqrt{1-2\epsilon}+1)$, for $\epsilon\rightarrow 0$ the
system and reservoir are decoupled while for $\epsilon\rightarrow 0.5$
the system and reservoir are strongly coupled.  The above S-Matrix
satisfies the conservation of current\cite{shapiro}, and accounts for
the possibility of strong to weakly coupled reservoirs through the
coupler ``$\epsilon$''. One of the important conclusions of the work
was that the magnitude of persistent current flowing in the loop
decreases with increasing coupling strength $\epsilon$ but without any
change in their nature, i.e., diamagnetic or paramagnetic. This effect
is solely due to exchange of carriers between reservoir and ring
(dephasing), and also this is true if the lead connecting reservoir to
ring has a charging energy much less than the level spacing.
Experimentally persistent currents in both open and closed systems
have been observed\cite{webb,levy,chandra,maily} and these
observations have given rise to a spurt in theoretical
activities\cite{rmp_persis}.

\begin{figure} [h]
\protect\centerline{\epsfxsize=3.5in \epsfbox{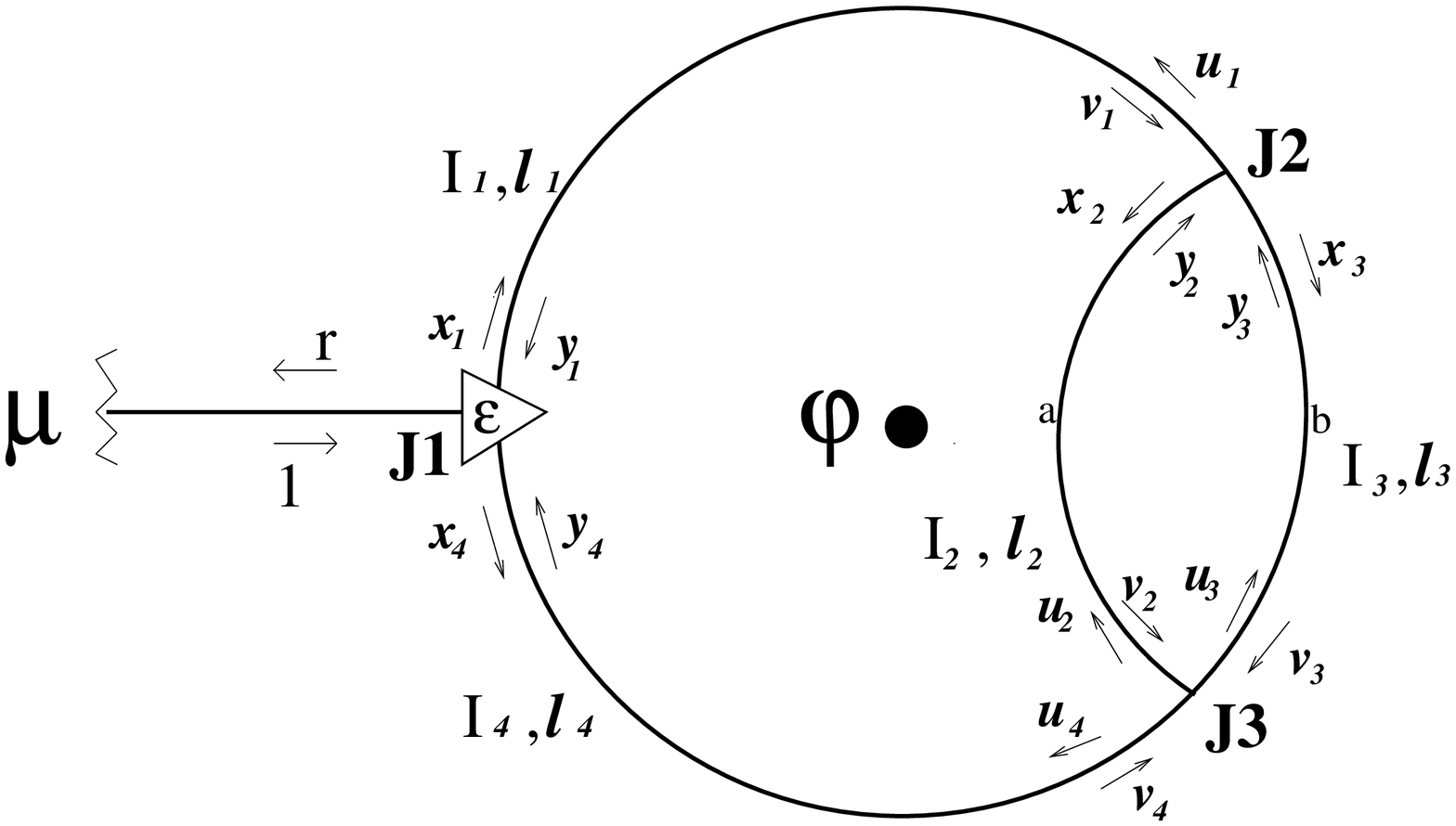}}
\caption{One dimensional mesoscopic ring coupled to a bubble with a
  lead connected to a reservoir at chemical potential $\mu$. The
  localized flux $\phi$ penetrates the ring.}
\end{figure}

\section {Model}
 
Let us consider the double ring system as shown in FIG.~1.  The static
localized flux piercing the loop is necessary to break the time
reversal symmetry and induce a persistent current in the system.  The
geometry we consider is a one-dimensional ring with an attached bubble
and a lead connected to a reservoir at chemical potential $\mu$. The
reservoir acts as an inelastic scatterer and as a source of energy
dissipation. All the scattering processes in the leads including the
loop are assumed to be elastic. Hence there is complete spatial
separation between the elastic and inelastic processes. The loops
J1J2aJ3J1 and J1J2bJ3J1 enclose the localized flux $\Phi$. However,
the bubble J2aJ3bJ2 does not enclose the flux $\Phi$. The same
S-Matrix coupler couples the double ring to the reservoir, i.e.,
$S_{J1}=S_J$, for the other two junctions we take symmetric
couplers\cite{porod}.

\[S_{J2}=S_{J3}=\left(\begin{array}{ccc}
-\frac{1}{3}    &  \frac{2}{3} &  \frac{2}{3}\\
 \frac{2}{3}   & -\frac{1}{3} &  \frac{2}{3}\\
  \frac{2}{3}     &   \frac{2}{3} &  -\frac{1}{3} 
\end{array} \right) \]

The waves in the four arms of the system depicted in FIG.~1 are
related as follows: The waves incident into the branches of the double
ring system are related by the S-Matrices for J1J2 arm by-

\[\left(\begin{array}{c}
y_1\\
v_1\\
\end{array} \right) \ =\left(\begin{array}{cc}
0     & e^{ikl_1} e^{-i \alpha_1}\\
e^{ikl_1} e^{i \alpha_1} & 0 \\
\end{array} \right) \left(\begin{array}{c}
x_1\\
u_1
\end{array} \right)\]
for J2bJ3 arm by-

\[\left(\begin{array}{c}
y_2\\
v_2\\
\end{array} \right) \ =\left(\begin{array}{cc}
0     & e^{ikl_2} e^{-i \alpha_2} \\
e^{ikl_2} e^{i \alpha_2} & 0 \\
\end{array} \right) \left(\begin{array}{c}
x_2\\
u_2
\end{array} \right)\]
 
for J2aJ3 arm by-

\[\left(\begin{array}{c}
y_3\\
v_3\\
\end{array} \right) \ =\left(\begin{array}{cc}
0     & e^{ikl_3} e^{-i \alpha_3} \\
e^{ikl_3} e^{i \alpha_3} & 0 \\
\end{array} \right) \left(\begin{array}{c}
x_3\\
u_3
\end{array} \right)\]

for J3J1 arm by-

\[\left(\begin{array}{c}
y_4\\
v_4\\
\end{array} \right) \ =\left(\begin{array}{cc}
0     & e^{ikl_4} e^{i \alpha_4} \\
e^{ikl_4} e^{-i \alpha_4} & 0 \\
\end{array} \right) \left(\begin{array}{c}
x_4\\
u_4
\end{array} \right)\]

Here $kl_1$, $kl_2$, $kl_3$ and $kl_4$ are the phase increments of the
wave function in the absence of flux.  $\alpha_1$, $\alpha_2
$,$\alpha_3$, and $\alpha_4$ are the phase shifts due to flux in the
arms of the considered double ring system.  Clearly,
$\alpha_{1}+\alpha_{2}+\alpha_{4}=\frac{2\pi\Phi}{\Phi_0} $, where
$\Phi$ is the flux piercing the loop and $\Phi_0$ is the flux quantum
$\frac{hc}{e}$, also
$\alpha_{1}+\alpha_{3}+\alpha_{4}=\frac{2\pi\Phi}{\Phi_0}$, thus
$\alpha_{2}=\alpha_{3}$\cite{colin_prb,colin_ijmpb}.  The current
densities (in dimensionless form) in the various arms of the system
are given as follows\cite{buti,colin_prb}- $I_{1}=|x_1|^2-|y_1|^2$,
$I_{2}=|x_2|^2-|y_2|^2$, $I_{3}=|x_3|^2-|y_3|^2$,
$I_{4}=-|x_4|^2+|y_4|^2$, wherein complex amplitude for propagating
waves $x_{1}, x_{2}, x_{3}, x_{4}, y_{1}, y_{2}, y_{3}, y_{4}, $ are
as depicted in Fig.~1. The induced current densities in the various
arms of the loop are assigned labels $I_{1}, I_{2}, I_{3}$ and $I_4$,
while the arm lengths of various parts of the ring are $l_{1}, l_{2},
l_{3}$ and $l_{4}$.

Now a question may arise as to why we consider the persistent current
densities in this special geometry of a double ring. This geometry
exhibits the current magnification effect in
equilibrium\cite{colin_prb} which is a quantum phenomena. The current
magnification effect, a purely quantum phenomena arises in equilibrium
and non-equilibrium systems\cite{mosk,deo_cm,colin_ijmpb} and leads to
large orbital magnetic moments. A double ring
system\cite{pareek_double} which does not exhibit current
magnification effect will show {\bf no disagreement} between the
magnetic moments calculated from the local current densities (via
Amperes law) and that from the derivative of the free energy, which we
have separately verified\cite{cb_unpub}.

\begin{figure*} [t]
\protect\centerline{\epsfxsize=7.0in \epsfbox{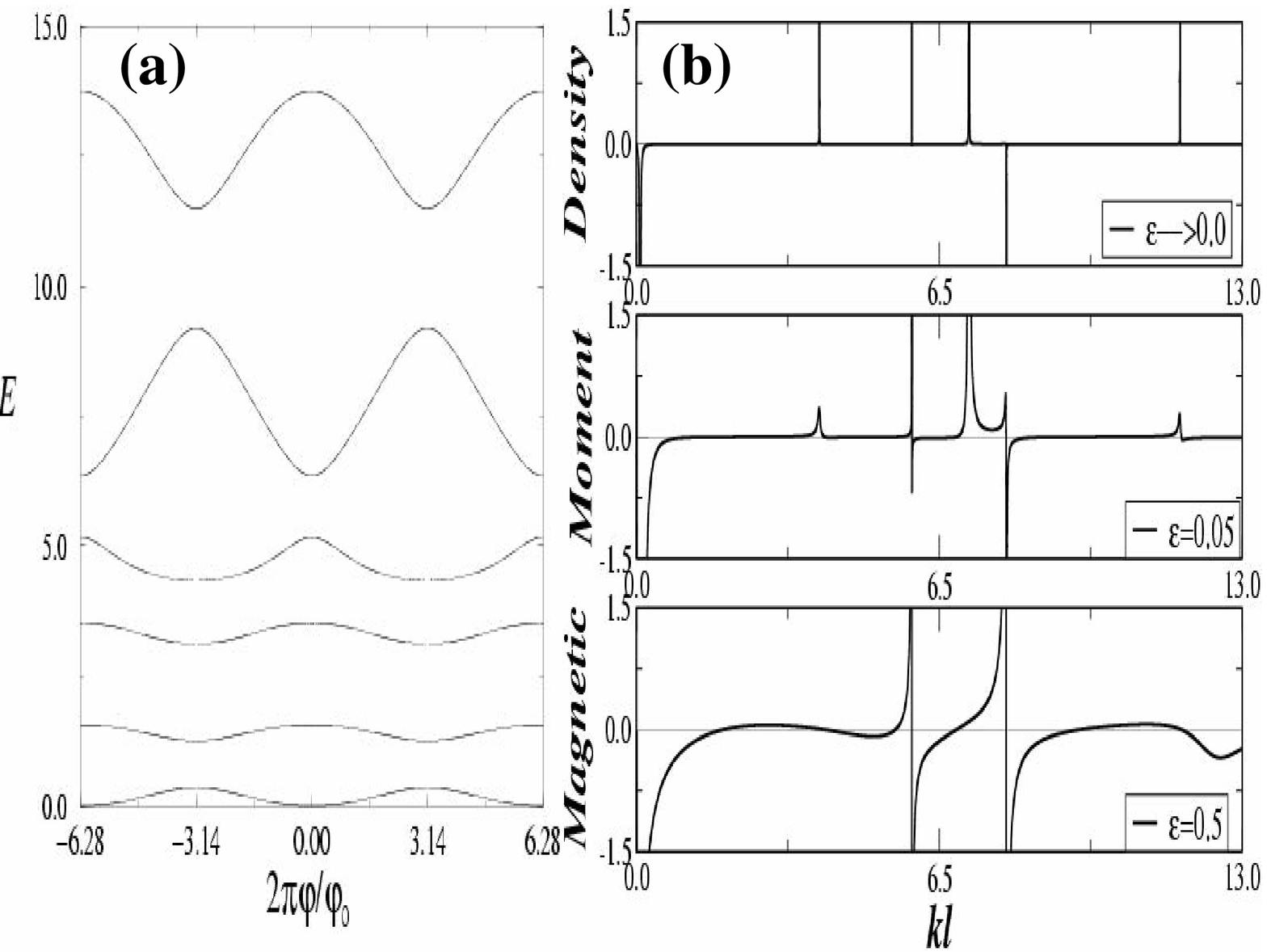}}
\caption{Plot of (a) energy levels and (b) orbital magnetic moment
  density $\mu_2$, for length parameters $l_{1}/l=l_{4}/l=0.375$, $l_{2}/l=
  0.15$, $l_{3}/l=0.85$ and flux $=0.1$. In (b) the upper, middle and lower panels are
  for coupling strengths $\epsilon\rightarrow0, \epsilon=0.05,$ and
  $\epsilon=0.5$ }
\end{figure*}

\begin{figure*} [!]
\protect\centerline{\epsfxsize=7.0in \epsfbox{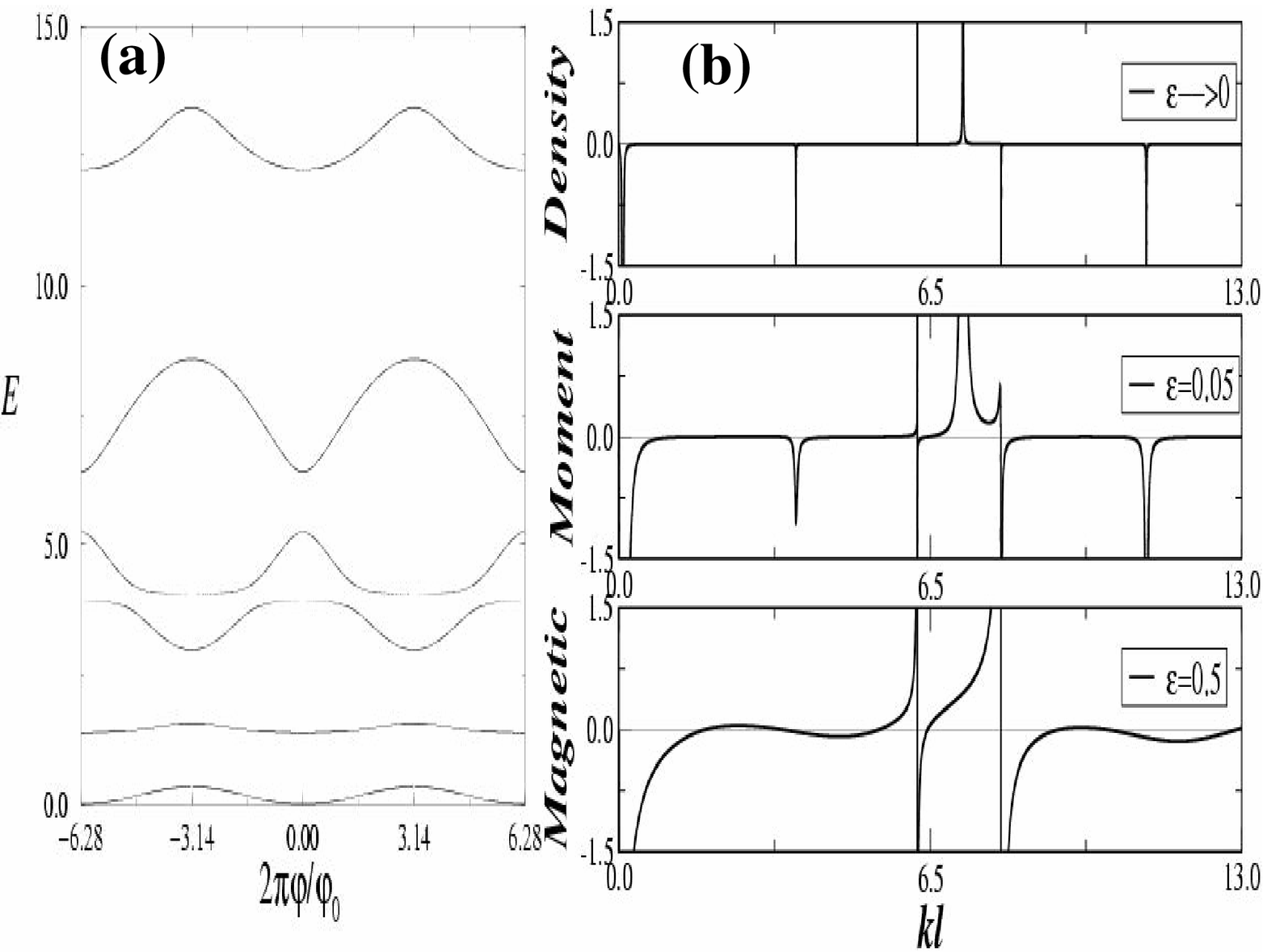}}
\caption{Plot of (a) energy levels and (b) orbital magnetic moment
  density $\mu_2$, for length parameters $l_{1}/l=l_{4}/l=0.375$, $l_{2}/l=
  0.05$, $l_{3}/l=0.95$ and flux $=0.1$. In (b) the upper, middle and lower panels are
  for coupling strengths $\epsilon\rightarrow0, \epsilon=0.05,$ and
  $\epsilon=0.5$ }
\end{figure*}

\section {Theory} 

We know that magnetic moment is the derivative of free energy with
respect to magnetic field, $\mu=-\frac{1}{c}\frac{\partial F}{\partial
  H}$, wherein $F$ is the free energy and $H$ is the magnetic field
enclosed by the system.  From text books we know that a current
carrying loop behaves as a magnet, in other words Amperes law which
states- magnetic moment is current multiplied by the area enclosed.
The orbital magnetic moment density for a system coupled to reservoir
can be calculated in two ways, first from the formulation of
Akkermans, et al, \cite{akker,mello} one can calculate it as follows-

\begin{eqnarray}
d\mu=\frac{1}{2\pi i}\frac {\partial[ln det S]}{\partial \phi} dE
\end{eqnarray}

Here, $d\mu$ is the differential contribution to the magnetic moment
at energy E, and S is the on-shell scattering matrix. In the system
considered in Fig.~1, the on-shell scattering matrix is just the
complex reflection amplitude $r$. Also from Amperes law one can
calculate the orbital magnetic moment density via the local currents
in the system. The orbital magnetic moment density defined via local
currents in a loop, depends strongly on the topology of the system,
whereas the magnetic moment densities calculated from the eigen
spectrum as also in the formulation of Akkermans, et al, do not.  This
is special to only one-dimensional geometry where the eigen-spectrum
is independent of variation in topology of the system . In fact there
are infinitely many topological structures
possible\cite{cedraschi,colin_ijmpb}. If we consider our system as
depicted in Fig.~1, to be planar and lying in the x-y plane then the
magnetic moment density ($\mu_1$) can be viewed as being generated by
current density $I_1$ enclosing an area $A_r$ and by current density
$I_3$ enclosing area $A_b$, i.e.,
$\mu_1=\frac{1}{c}(I_{1}A_{r}+I_{3}A_{b})$, wherein $A_{r}$ and
$A_{b}$ are the areas enclosed by the ring ($J1J2aJ3J1$) and the
bubble ($J2aJ3bJ2$) respectively. Another orientation of the system in
which the arm $J2bJ3$ is in the x-z plane gives
$\mu_2=\frac{1}{c}(I_{1}A_{r}+I_{2}A_{r})/2=I_c A_r/2$ wherein
$I_{c}=I_{1}+I_{2}$ is said to be the most appropriate generalization
of the equilibrium persistent current and which is
consistent\cite{cb_unpub} with Eq.~1, see [\onlinecite{cedraschi}] for
further details.  Several other orientations are possible, for
example, if the bubble lies in x-y plane and the ring lies in x-z
plane, then $\mu_{z}=\frac{1}{c}(I_{3}A_{r}-I_{2}A_{r})/2$ and
$\mu_{y}=\frac{1}{c}I_{1}A_{r}$. Even when our system lies in the x-y
plane for fixed $l_1, l_2, l_3$, by deforming their shapes we can have
different values of magnetic moment density along the z direction.  It
is also worth mentioning that the total magnetic moment (at
temperature $T=0$) of a representative system is obtained by
integrating the magnetic moment densities up-to the Fermi wave-vector
$k_f$.

\section  {Results} 
From the Amperes law we calculate the orbital magnetic moment density
for some length parameters. After calculating we plot the orbital
magnetic moment densities obtained for the case of
$\mu_{2}=\frac{1}{c}(I_{1}A_{r}+I_{2}A_{r})/2=I_c A_r/2$. In Fig.~2(a)
we depict the plot of the energy-eigen values (normalised by $\pi^2$) of the closed system,
and in Fig.~2(b) we plot the dimensionless orbital magnetic moment
density $\mu_2$ obtained via the local persistent current densities as
a function of the dimensionless Fermi wave-vector $kl$ for different
coupling parameters. Now lets test the equivalence of these
definitions as $\epsilon \rightarrow 0$, i.e., the reservoir and
system are almost decoupled. We see complete agreement. It can be
noted from Fig.~2(a) that the ground state carries diamagnetic current
while 1st, 2nd and 3rd excited states carry paramagnetic current for
small values of flux (which is obvious from their slopes), the 4th
excited state carries diamagnetic currents while the fifth
paramagnetic current. Similarly from the top most panel of Fig.~2(b),
the $\epsilon \rightarrow 0$ limit, it can be seen that the ground
state carries diamagnetic current (magnitude is negative) while 1st,
2nd and 3rd excited states carry paramagnetic currents (magnitude is
positive), the 4th carries diamagnetic current while the fifth
paramagnetic current. Indeed the two definitions are completely
equivalent as far as the $\epsilon \rightarrow 0$ limit (henceforth
referred to as the weak coupling case) is considered.  Now as we
increase $\epsilon$, we notice a dramatic change from the weak
coupling case (see middle panel in FIG.~2(b)). There are
paramagnetic-diamagnetic jumps at those Fermi energies wherein one
would have expected pure diamagnetic or paramagnetic current. This
behavior is more seen as one increases the coupling till the maximum
($\epsilon=0.5$) is reached, see lowest panel of Fig.~2(b). Looking
closely at the middle panel of Fig. 2(b), one can notice that the
peaks broaden, noticeably the ground, 1st, 3rd and 5th, while the
nature of the currents carried at 2nd and 4th levels {\bf changes
  qualitatively}. Also, as one approaches the $\epsilon=0.5$ limit the
1st, 3rd and 5th levels disappear completely, these are some of the
other qualitative features which distinguish the cases depicted in
Fig. 2(a) and (b). This reaffirms that predictions from equilibrium
thermodynamics are in general not valid for a quantum system strongly
coupled to a bath. In the conventional equilibrium treatment the
properties of the bath appear only through a single parameter namely,
the temperature $T$ and nowhere the coupling parameter appears in the
magnitude of equilibrium physical quantities. Our results reaffirm
that equilibrium properties are determined by the coupling parameter
in the quantum domain.  The finite coupling can lead to qualitative
changes (and not just the broadening of levels or persistent current
peaks) from that of the predictions of equilibrium statistical
mechanics as shown above.

We have verified separately that the orbital magnetic moment density
calculated by the formulation of Akkermans, et al, (Eq.~1) is similar
to that in Fig.~2(b), for the geometry considered here\cite{cb_unpub}.
In addition we also have calculated the orbital magnetic moment
density for different topological configurations, e.g.,
$\mu_{1},\mu_{z}, \mu_{y}$ (see the {\em Theory} section for further
details) and obtained results not in consonance with that calculated
from the eigen-spectrum (equilibrium statistical mechanics), thus
bolstering the fact that the orbital magnetic moment density
calculated from the local current densities is inherently linked to
the topology of the system. As already pointed out, the orbital
magnetic moment density calculated through the local current densities
is qualitatively different (nature) from that calculated from the
closed system energy eigen-spectrum which is independent of
topological variations.

We have seen these novel features not only for one set of length
parameters but for many different set of length parameters. In
Fig.~3(a), we plot the energy eigenvalues (normalised by $\pi^2$) as a function of flux, and
in Fig.~3(b) the dimensionless orbital magnetic moment densities as a
function of the dimensionless Fermi wave-vector $kl$ for different
length parameters. Herein also it can be seen from Fig.~3(a) that the
ground and the first excited states carry diamagnetic currents while
2nd and 3rd carry paramagnetic currents for small values of flux
(which is obvious from their slopes), the 4th and 5th again carry
diamagnetic currents.  Similarly from the top most panel of Fig.~3(b),
the $\epsilon \rightarrow 0$ limit, it can be seen that the ground and
first excited state carry diamagnetic current (magnitude is negative)
while 2nd and 3rd carry paramagnetic currents (magnitude is positive),
the 4th and 5th carry diamagnetic currents. The two definitions are
again completely equivalent as far as the $\epsilon \rightarrow 0$
limit is considered. Again as we increase $\epsilon$, we notice the
same change from the weak coupling case (see middle panel in
FIG.~3(b)). There are paramagnetic-diamagnetic jumps at those Fermi
energies wherein one would have expected pure diamagnetic or
paramagnetic current. This behavior is more seen as one increases the
coupling till the maximum ($\epsilon=0.5$) is reached, see lowest
panel of Fig.~3(b). Thus, in contrast, to the case of a single quantum
ring coupled to a reservoir as was considered by B\"{u}ttiker, wherein
coupling only led to a broadening of energy levels\cite{buti}, in the
case of a quantum double ring system considered here, in addition to
level broadening one also sees {\bf a change in nature} of currents as
one increases the strength of coupling to the reservoir.

One should also mention here that there is a drawback in modeling the
inelastic effects by this model, if one changes the position of
attachment of lead and the double ring system. Qualitatively,
different results for the orbital magnetic moment density would be
obtained as by definition it involves the length parameters of our
system, These length parameters will of-course change with position of
junction (J1) . Only in the very weakly coupled regime can the
specific position of lead to reservoir attachment be ignored.
Of-course this drawback does not exist for the system investigated by
B\"{u}ttiker in Ref.\onlinecite{buti} wherein the orbital magnetic
moment density would be same irrespective of the position where lead
is attached.  However, this simple model for coupling between system
and reservoir is enough in order to bring out the importance of finite
coupling between system and bath, {\it vis a vis} equilibrium
thermodynamics.

\section {Conclusions} 

In this work we have shown using a very simple model of
system-reservoir coupling that the equilibrium properties are
determined by the strength of coupling parameter. This is consistent
with the recent findings\cite{TA,dattgup_prl}, that in the quantum
domain, equilibrium properties of the sub-system are related to
dissipative coefficients arising due to subsystem-bath coupling,
unlike the predictions of conventional statistical mechanics. These
fully dynamical studies in the quantum domain invoke coherence and
entanglement between system and bath degrees of freedom to come to the
same conclusion. However our treatment is simple and invokes the
single particle coherence which is disrupted by the presence of the
reservoir (without bringing the notion of quantum entanglement between
system and bath). However these finite coupling induced qualitative
changes can be observed only in hybrid rings which exhibit current
magnification effect, a purely quantum phenomena, in
equilibrium\cite{colin_prb}. We have verified separately
that multiple ring structures which do not exhibit current
magnification effect do not show these qualitative changes apart from
features expected from broadening of energy levels\cite{cb_unpub}.  These results can
be verified experimentally by attaching a voltage probe coupled to a
nanoscopic semi-conducting system as depicted in Fig.~1. Here the
voltage probe serves the purpose of a reservoir. One can change the
parameters of the junction between system and lead by applying
appropriate gate voltage (to simulate the effects of coupling
parameter) beneath the junction and measure the orbital magnetic
response as function of the gate voltage.

\section {Acknowledgments}

 Authors thank Dr. P. S. Deo for useful discussions.

\end{document}